\documentclass[aps,prb,twocolumn,showpacs,amsmath,amssymb,superscriptaddress,floatfix,]{revtex4}
\usepackage{epsfig}
\usepackage[latin9]{inputenc}
\usepackage{color}
\usepackage{hyperref}
\usepackage{graphicx}
\usepackage{hypernat}

\def\etal{{\em{et al.}}}
\begin{document}


\title{First-Principle Wannier function analysis of the electronic structure of PdTe: 
Weaker magnetism and superconductivity}

\author{Chinedu E. Ekuma}
\altaffiliation{corresponding email: cekuma1@lsu.edu}
\affiliation{Department of Physics \& Astronomy Louisiana State University,
Baton Rouge, LA 70803, USA}
\affiliation{Center for Computation and Technology, Louisiana State University, Baton Rouge, LA 70803, USA}

\author{Chia-Hui Lin}
\affiliation{Department of Physics and Astronomy, Stony Brook University, Stony Brook, New York 11794, USA}
\affiliation{Condensed Matter Physics and Materials Science Department,
Brookhaven National Laboratory, Upton, New York 11973, USA}

\author{Juana Moreno}
\affiliation{Department of Physics \& Astronomy Louisiana State University,
Baton Rouge, LA 70803, USA}
\affiliation{Center for Computation and Technology, Louisiana State University, Baton Rouge, LA 70803, USA}

\author{Wei Ku}
\affiliation{Department of Physics and Astronomy, Stony Brook University, Stony Brook, New York 11794, USA}
\affiliation{Condensed Matter Physics and Materials Science Department,
Brookhaven National Laboratory, Upton, New York 11973, USA}

\author{Mark Jarrell}
\affiliation{Department of Physics \& Astronomy Louisiana State University,
Baton Rouge, LA 70803, USA}
\affiliation{Center for Computation and Technology, Louisiana State University, Baton Rouge, LA 70803, USA}

\date{\today}

\begin{abstract}
\noindent We report a first-principles Wannier function study of the electronic structure of 
PdTe. Its electronic structure is found to be a broad three-dimensional Fermi surface with highly reduced 
correlations effects. In addition, the higher filling of the Pd $d$-shell, its stronger 
covalency resulting from the closer energy of the Pd-$d$ and Te-$p$ shells, and the larger crystal field effects 
of the Pd ion due to its near octahedral coordination all serve to weaken significantly electronic correlations 
in the particle-hole  (spin, charge, and orbital) channel. In comparison to the Fe Chalcogenide e.g., FeSe, 
we highlight the essential features 
(quasi-two-dimensionality, proximity to half-filling, weaker covalency, and higher orbital degeneracy) 
of Fe-based high-temperature superconductors.

\end{abstract}

\pacs{31.15.A-, 74.70.Ad, 31.15.V-, 71.15.Ap,71.20.-b, 71.27.+a}

\maketitle

\section{Introduction}

The discovery of high-temperature superconductivity in the Fe chalcogenides \cite{Hsu2008,Ozaki2012} 
led to a ``gold rush'' to find superconductivity in other non-cuprate materials. Compared with 
Fe-based superconductors, \cite{PhysRevLett.104.187002,Kamihara2008,Wang2008,Norman2008,
PhysRevLett.101.107006, PhysRevB.85.014109,
Parker2009,PhysRevB.80.214508,PhysRevB.84.054528} these non-toxic layered compounds with weak 
inter-layer van der Waals forces exhibit interesting physical properties, 
including phase separation, \cite{Hu2011} strong correlations, \cite{PhysRevB.84.214407,Jinsheng2011} non-trivial 
isovalent doping, \cite{PhysRevB.78.224503,Gawryluk2011,PhysRevB.82.020508} 
Fe excess effects, \cite{Sudesh07E119} Fe vacancy, \cite{Berlijn,PhysRevX.1.021020,PhysRevB.83.140505}  
and rich high-pressure phase diagrams. 
\cite{PhysRevB.84.214407,Jinsheng2011,Mizuguchi2008,Mizuguchi2009,PhysRevB.80.064506} 
Attention has also shifted to other non-Fe transition metal based 
chalcogenides as they are shedding light on the road to the \emph{post-iron age} in superconductivity.

\begin{figure}[htb]
\begin{center}
  \includegraphics[width=1\columnwidth,clip=true]{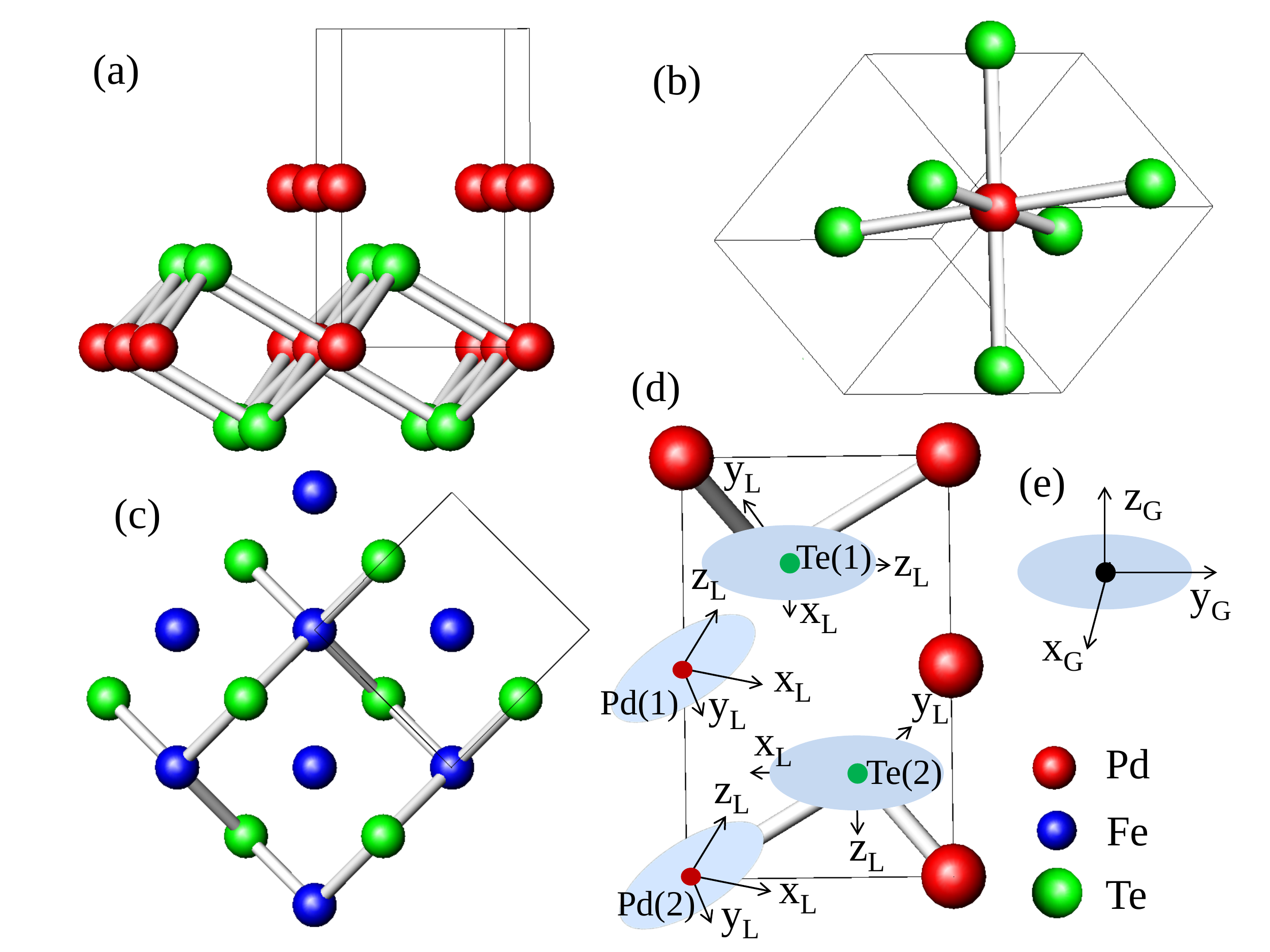}
 \caption{(Color online) (a) The hexagonal structure of PdTe along the [001] direction. 
The positions of the atoms are Pd(1): (0,0,1/2), Pd(2): (0,0,0), Te(1): (1/3,2/3/3/4), and Te(2): (2/3,1/3,1/4), respectively. 
(b) The Pd-centered edge and face sharing environment with the neighboring octahedra. There are six Te octahedrally 
arranged around Pd. (c) The tetragonal structure of FeTe along the [001] direction. (d) The local coordinate of Pd 
and Te in the PdTe system as utilized in the definition of the Wannier basis. The local coordinates corresponding to 
each of the atoms in the unit cell (two Pd and two Te) have been defined using the x-convention. (e) The global coordinate 
of the PdTe system. The  subscripts L and G denote local and global coordinates, respectively.}
\end{center}
 \label{fig:PdTe_struct}
 \end{figure}

There are yet no rigorous computational study of the electronic structure of Palladium monotelluride (PdTe). 
So far, experimental studies show that it has a $T_c\approx$ 
2.3 -- 4.5 K,\cite{Karki2012,PhysRev.92.874,PhysRev.90.487,Kjekshus1965}, electron-phonon coupling constant 
$\lambda_{e-p}$$\sim$ 1.4, \cite{Karki2012}  and a phase diagram reminiscent of 
the high-pressure diagram of Fe chalcogenides (FeCh). \cite{Mizuguchi2009,C2CS35021A} Compared with 
FeCh~\cite{Hsu2008,Ozaki2012,PhysRevB.81.094115,PhysRevB.78.224503,PhysRevB.80.064506} 
the hexagonal PdTe structure (c.f. Fig.~\ref{fig:PdTe_struct}(a)) can be regarded as the deformed structure of FeTe 
(c.f. Fig.~\ref{fig:PdTe_struct}(c)) obtained by sliding the anion and cation layers.
Despite such structural similarity, the local ligand field is transformed from a 
tetrahedral cage surrounding Fe to an octahedral cage surrounding Pd. With a similar crystal structure 
but distinct physical properties, PdTe thus serves as a good candidate for comparison
to better illuminate the important physical effects and to reveal the building blocks
for a stronger superconductivity in the Fe chalcogenides. 

Motivated by the growing effort to understand properties of NiAs-like systems due to their very 
diverse and interesting properties, we provide a first-principles 
Wannier function analysis of the electronic structure of PdTe. We remark 
that there is no yet `rigorous' electronic structure study of the properties of PdTe. 
A concomitant motivation 
is the evolving effort to understand the superconductivity mechanism in Fe-based superconductors. 
By comparing the characteristics of PdTe with those of Fe chalcogenides, one expects 
to identify key ingredients behind the high-temperature superconductivity in 
Fe-based superconductors, a current research topic of general interest to a broad audience. 

We find that the 
face-shared octahedral coordination of Pd and the larger size of its 4$d$ orbital favor the electronic kinetic
energy over the Coulomb potential energy. Also, a broad three-dimensional Fermi surface (FS) 
with a strong variation along the $k_z$ direction, and a lack of orbital degeneracy are observed. 
The almost fully occupied Pd $d$ orbitals lead to a strong covalent Pd-Te bonding and a vanishing local magnetic moment. 
A near octahedral coordination in PdTe leads to a large crystal field 
splitting of the Pd-$d$ orbitals ($t_{2g}$ -- $e_g$) with a crystal field parameter of $\Delta_{oct}$ $\sim$ 400 meV.  
Therefore, most probably electronic correlations are suppressed in pure PdTe. 
One can thus conclude that due to these dilute correlations, 
pure PdTe may just be a low $T_c$ superconductor in basic agreement with the experimental work of Karki \etal 
\onlinecite{Karki2012}. 
This study illuminates the important underlying physics of the high-Tc Fe-based superconductors and most 
especially, reveals the building blocks
for a stronger superconductivity in Fe chalcogenides. Such a study can help to reassess the rules one uses 
in the search of a room temperature superconductor and 
in particular, providing key insights on how to achieve high temperature
superconductivity in general.

\section{approach and crystal structure}
We use first-principles calculations to study the electronic structure of PdTe and the competition between 
various magnetic configurations. We utilize standard density functional theory (DFT) within the general potential 
linearized augmented planewave (LAPW) method \cite{singh2006} and  the generalized gradient approximation 
(PBE-GGA) functional, \cite{PhysRevLett.77.3865} as implemented in WiEN2k. \cite{Blaha2001} 
In our computations, we utilize the 
room temperature experimental lattice parameters: a = b = 4.152 \AA{} and c = 5.672 \AA{}, \cite{Gronvold1956} 
and the hexagonal crystal structure with space group P6$_3$/mmc 
(Patterson symbol). \cite{Hahn2005}
Pd and Te atoms occupy $2a$ and $2c$ Wyckoff positions, respectively, namely, Pd(1): (0,0,1/2),
Pd(2): (0,0,0); Te(1): (1/3,2/3/3/4), and Te(2): (2/3,1/3,1/4). \cite{Hahn2005,Fluck1996,Kjekshus1965} 
Thus, the first-principles ground 
state can be reached with all the WiEN2k default settings and a k-mesh of $14\times 14\times 9$. 
To downfold the DFT electronic band structure, symmetry respecting Wannier functions
\cite{PhysRevLett.89.167204,PhysRevLett.104.216401,PhysRevLett.106.077005}
of Pd $d$ and Te $p$ are constructed to capture the low-energy Hilbert space
within [-8, 3] eV and obtain an effective tight-binding Hamiltonian. We then 
obtain the band structure and the Fermi surface by calculating 
the orbital-resolved spectral function, $A_{n,k}(\omega)$, where $n$, $k$, $\omega$ are 
orbital index, crystal momentum, and energy, respectively.

The structure of PdTe is tied strongly to its electronic and magnetic
properties, since the spatial extension of the 4$d$ orbitals 
promotes comparable and competing kinetic and Coulomb energies. 
PdTe crystallizes in the  NiAs or D$_{6h}^4$ crystal structure, which is of interest both from 
experimental and theoretical points of view.\cite{K1978,Sandratskii1981}
Transition metal compounds with the NiAs crystal structure attract special 
interest due to anomalies in their magnetic, elastic, and electrical properties, especially near phase 
transitions, the nature of which is still under study.\cite{Sandratskii1981}
In PdTe the Pd atoms sit in an fcc-like environment, while the Te atoms form an hcp-like structure and 
are surrounded by six Pd metal atoms forming a trigonal prism. \cite{Grundmann2006} 
In this structure, the Pd-centered octahedron shares both edges and faces with the neighboring PdTe$_6$ octahedra 
(c.f. Fig.~\ref{fig:PdTe_struct} (a) and (b)).
This is different from FeCh or FePn where the corrugated square planes of Fe with Se (or As)  
atoms are such that Fe is tetrahedrally coordinated below and above the planes. Thus, the Fe atoms
sit in the tetrahedral cavities of the tetragonally distorted close packed lattice of Se (or As).\cite{Singh2009} 
We also note that the structure of PdTe places Pd atoms closer to each other than in a perovskite; as such, direct 
Pd-Pd hybridization becomes important as we will discuss below.

\section{Results and discussion}

\begin{figure}
\begin{center}
 \includegraphics[trim = 4mm 9mm 0mm 10mm,width=1\columnwidth,clip=true]{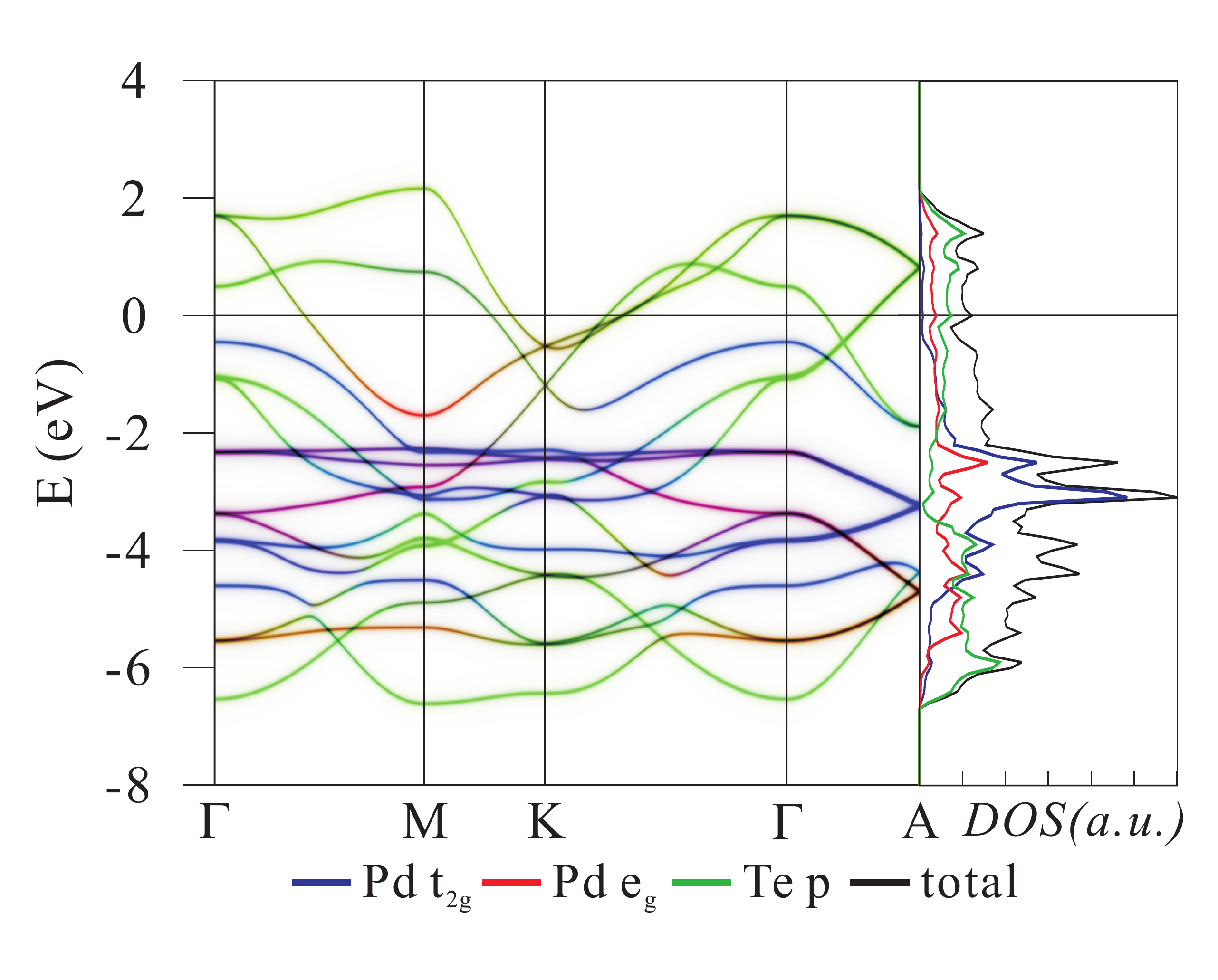}
\caption{(Color online) Left panel: Calculated dispersion (band structure)
of PdTe within [-8.0, 3.0] eV. Right panel: The corresponding spectra (in a.u. unit)
as obtained from the symmetry
respecting Wannier functions within [-8.0, 3.0] eV. In both plots, the Fermi level is set equal to zero.}
\label{fig:PdTe_elect}
\end{center}
\end{figure}

Our resulting electronic band structure and density of states (DOS) for non-magnetic PdTe are shown in 
Fig.~\ref{fig:PdTe_elect}, colored to emphasize the Pd $t_{2g}$, $e_g$ and Te $p$ orbitals. The whole spectrum is 
predominantly a single huge band complex due to the strong hybridization between Pd $d$ and Te $p$ orbitals. The 
role of $p-d$ hybridization in NiAs-type compounds have been highlighted in literature (See for 
e.g., Refs. \onlinecite{PhysRevB.10.995,PhysRevB.61.16370,PhysRevB.17.1828} for discussions). Notice from the DOS 
(c.f.~Fig.~\ref{fig:PdTe_elect}) that almost all the Pd $d$ orbitals
are occupied, and the low energy excitations near the Fermi surface are composed mainly of Te $p$ electrons
hybridized with the $e_g$ orbital of Pd. The Pd $d$ orbital occupancy can 
be calculated from the trace of the one-particle reduced density matrix on each Pd atom 
(in Wannier function basis) and is found to be 9.31, which is very close to the fully 
occupied value of 10. This indicates strong covalency of the Pd-Te bonding and strong suppression 
of the spin local moment. The strong covalency can also be confirmed by measuring the distance 
$\Delta E_{p-d}$ between the highest peaks in Te 5$p$ and Pd 4$d$ bands which is
a measure of the ionicity or covalency of the bonding. $\Delta E_{p-d}$ is found to be $\approx$ 2.92 a.u. 
We find in particular that the electronic structure of PdTe 
is determined by short-range interactions in the Te $p$ -- Pd $d$ band complex, with the 
ligand-field splitting of the Pd $d$ states in the 
environment of the Te atoms being the key determinant of the structure of the $d$ band.

\begin{figure}
\begin{center}
  \includegraphics[width=1.0\columnwidth,clip=true]{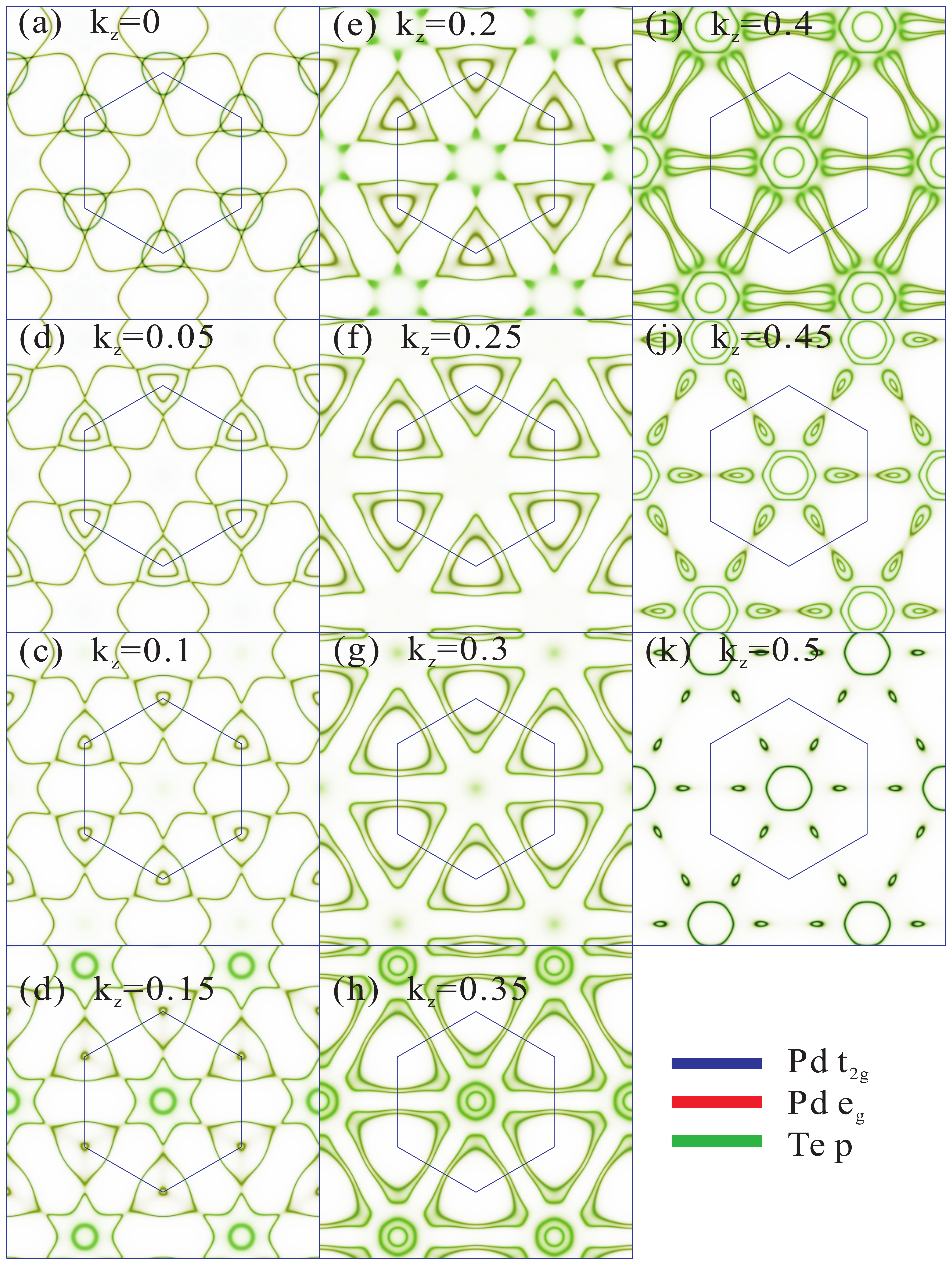}
 \caption{(Color online) The Fermi surface of PdTe at different $k_z$ 
planes with the same color code as Fig.~\ref{fig:PdTe_elect}. The value of $k_z$ is in units of $2 \pi /c$. 
The states in the proximity of the Fermi surface are predominantly Te-$p$ states, hence the color of the Fermi surface 
plots is mainly green.}
 \label{fig:fig3}
\end{center}
 \end{figure}

In Fig.~\ref{fig:fig3}, the Fermi surface is presented in various $k_z$ planes. As already seen 
in the band structure, the states in the proximity of the Fermi level consist mainly of Te $p$ 
(green orbital) weakly hybridized with Pd $e_g$ (red orbital), thus appearing dark green. 
The Fermi energy is located close to the local maxima of the DOS, corresponding to N(E$_\mathrm{F}$) $\approx$ 2.41 a.u. 
The corresponding electronic specific heat coefficient 
$\gamma$ is 2.84 mJ mol$^{-1}$ K$^{-2}$. This value is in good agreement with the experimental value of 6.0 
mJ mol$^{-1}$ K$^{-2}$ obtained by Karki \etal \onlinecite{Karki2012} if we take into account the 
many-body effect contributions from electron-phonon interactions such that $\gamma$ is renormalized as $\gamma_n$ 
$=$ $\gamma (1 + \lambda_{e-p})$.

To investigate the magnetism of PdTe, we survey various magnetic configurations, 
including collinear, bicollinear, and checkerboard antiferromagnetism, with spin-polarized 
calculations. Indeed, the non-polarized case possesses the lowest ground state total energy, 
so there are no long-range magnetic correlations. This is consistent with the fact 
that the almost-full-Pd-$d$ orbitals not only rule out the superexchange 
mechanism for antiferromagnetic correlations, but also suppress the effectiveness 
of Hund's coupling, and thus ferromagnetic correlations as well.

In order to gain a microscopic insight into the electronic structure of PdTe, we transform the 
self-consistent Kohn-Sham (DFT) Hamiltonian to Wannier basis, as summarized in Table~\ref{table:PdTe_Ham}.
In the intra-atomic Pd block, one finds a large ($\sim$400 meV) $t_{2g}$-$e_g$ splitting
which originated from the octahedral ligand field, with negligible off-diagonal terms associated with
the tiny distortion of the octahedral cage.
A similar splitting ($\sim$500meV) is also found between Te $p_z$ and $p_x/p_y$.
Furthermore, we find considerable inter-atomic Pd-Te and Pd-Pd hopping, 
for example, $t_{Pd(1)_{z^2},Te(1)_{p_x}} = 509$, $t_{Pd(1)_{z^2},Te(1)_{p_y}} = 881$,  
 $t_{Pd(1)_{x^2-y^2},Te(2)_{p_x}} = -881$, $t_{Pd(1)_{z^2},Te(1)_{p_z}} = 559$, 
and $t_{Pd(1)_{xy},Pd(2)_{d_{xy}}} = -353$ meV, 
which are much larger than the intrinsic scale for the on-site energy difference between Pd and Te ($\sim$100 meV) 
(c.f. Table~\ref{table:PdTe_Ham}). The same hopping value of 
$t_{Pd(1)_{z^2},Te(1)_{p_y}}$ and $t_{Pd(1)_{x^2-y^2},Te(2)_{p_x}}$, even when they are not symmetry related, 
is due to the 30$^o$ tilting angle that is naturally built-in in the crystal symmetry of PdTe, as can be understood from 
the Slater-Koster coefficients used in the tight-binding model.
\cite{PhysRev.94.1498,Harrison1989}

\begin{table}[htb]
\caption{(Color online). This table is mainly intended to show the hopping dominated physics in PdTe 
and, as such, it represents a reduced matrix of the original 16 by 16 matrix of Pd $d$ and Te $p$ states. 
For clarity, the Pd(1) $d$ and Te(1) $p$ sub-bands are
represented using bold letters. The expectation value 
$\langle$$\mathrm{WF_i}$$\rvert$$\mathrm{H}$$\rvert$$\mathrm{WF_j}$$\rangle$, where $\mathrm{WF_i}$ 
corresponds to the Wannier function of orbital $i$, 
is denoted as $\langle$$\mathrm{n}$$\rvert$$\mathrm{H}$$\rvert$$\mathrm{n^\prime}$$\rangle$ on the table.
Onsite energies of Pd $d$ orbitals, split due to crystal fields 
into e$_\mathrm{g}$ ($z^2$ and x$^2$ -- y$^2$) (red) and t$_\mathrm{2g}$ 
(yz, xz, and xy) (blue) sub-bands, and the Te $p_x$, $p_y$, and $p_z$ sub-bands (green), 
and the hopping integrals among Pd $d$ and Te $p$ 
Wannier orbitals for the nonmagnetic case are displayed.
Units are meV. The local coordinates of each of the atoms are defined in Fig.~\ref{fig:PdTe_struct}(d).}
\begin{tabular}{ p{1.2cm}ccccc|cccc }
\hline\hline                        
$\langle$$\mathrm{n}$$\rvert$$\mathrm{H}$$\rvert$$\mathrm{n^\prime}$$\rangle$  &$\mathbf{z^2}$ & $\mathbf{x^2-y^2}$ &
$\mathbf{yz}$ & $\mathbf{xz}$ &$\mathbf{xy}$  &$\mathbf{x}$  &$\mathbf{y}$  &$\mathbf{z}$ \\ [0.5ex]   
\hline
$\mathbf{z^2}$ & \textcolor{red}{-2608} & 0 & 2 & 2 & 4  &509 & 881 & 559\\
$\mathbf{x^2-y^2}$  & 0 & \textcolor{red}{-2608} & -4 & 4 & 0 & 38 & -22 & 0 \\
$\mathbf{yz}$ & 2 & -4 & \textcolor{blue}{-2987} & -42 & 42  & 187 & -290 & 361\\
$\mathbf{xz}$ & 2 & 4 & -42 & \textcolor{blue}{-2987} & 42 & -345 & 17 & 361\\
$\mathbf{xy}$ & 4 & 0 & 42 & 42 & \textcolor{blue}{-2987} & 26 & 45 & 33\\
\hline
$\mathbf{x}$ & 509 & 38 & 187 & -345 & 26  & \textcolor{green}{-3112} & 0 & 0\\
$\mathbf{y}$ & 881 & -22 & -290 & 17 & 45 & 0 & \textcolor{green}{-3112} & 0 \\
$\mathbf{z}$ & 559 & 0 & 361 & 361 & 33 & 0 & 0 & \textcolor{green}{-2597} \\
\hline\hline
$z^2$ & -20 & 0 & 60 & 60 & 120  & 8 & 14 & 50 \\
$x^2$ -- $y^2$ & 0 & 20 & 104 & -104 & 0 & -22 & 13 & 0\\
$yz$ & 60 & 104 & -150 & -353 & 150 & 2 & 0 & -33\\
$xz$ & 60 & -104 & -353 & -150 & 150 & -1 & 1 & -33\\
$xy$ & 120 & 0 & 150 & 150 & -353  & 18 & 31 & -1\\
\hline
$x$ & 509 & -881 & 52 & 158 & -158  & 79 & 254 & 263 \\
$y$ & -38 & -22 & 0 & 307 & 307 & 254 & -214 & -455 \\
$z$ & 279 & -484 & 33 & -361 & 361  & -263 & 455 & 524\\
\hline\hline
\end{tabular}
\label{table:PdTe_Ham}
\end{table}

Unconventional superconductivity generally emerges in frustrated systems 
\cite{Scalapino1999,PhysRevLett.101.097010,Yang2007,PhysRevB.82.165118,Lacroix} when
long-range order in the spin, charge or orbital channel is suppressed.  The key ingredients needed for 
a typical unconventional superconductor, strong short-range 
correlations in these channels, are either negligible or dilute in PdTe. 

Several factors make PdTe very different from FeCh. Strong hopping due 
to the more extended 4$d$ orbitals, which promotes a larger kinetic energy, closer Pd-Pd distance in the face-shared 
Pd octahedron environment, negligible Hund's moment, weaker magnetic spin fluctuations, lack of orbital degeneracy, 
strong covalency, and very dilute magnetic frustration. Although PdTe has a layered structure similar to  the
tetragonal Fe chalcogenides, its intrinsic Pd $4d$ character and face-shared 
octahedron environment not only suppress the electronic correlations in the charge, orbital, 
and spin channels but also promote three dimensionality. 
The three dimensionality of PdTe  electronic structure is made clear from the 
strong $k_z$ variation of the Fermi surface topology in Fig.~\ref{fig:fig3}. 
Such a strong three-dimensionality indicates a much weaker Fermi surface nesting, and would likely 
favor a weaker long-range correlation in the particle-hole  (spin, charge, and orbital) channel. 
In addition, since the charge-transfer energy between Pd $d$ and chalcogen $p$ states is 
much smaller in PdTe than in FeCh, the Pd $d$ states are almost fully occupied, and 
there are no states available to form a large Hund's moment, making magnetic correlations
and, e.g., magnetic frustration effects, weaker.
Therefore, any possibility of magnetically-driven superconductor is suppressed. 
The relatively low superconducting temperature 
in PdTe may likely be due to another source, presumably phonons.

As can be seen from Table~\ref{table:PdTe_Ham}, the dominant hoppings are those between Pd and the neighboring Te. 
This strong hopping is due to the large orbital overlap between  Te and Pd ions 
(c.f. Fig.~\ref{fig:PdTe_struct}(d)). On the other hand, a strong Pd-Pd hopping is also observed. 
This is due to the spatial extension of the 4$d$ orbitals, a consequence of its larger quantum number 
compared to 3$d$ orbitals, that overlap with neighboring Pd ions; and the 
closer Pd-Pd distance in the octahedral environment.
Since the hybridization matrix between the $d$-orbitals scales as d$^{-4}$, where d is the bond length, 
a shorter bond length  greatly affects the hybridization between $d$-orbitals.\cite{PhysRevB.62.11639}  
This will increase the kinetic energy,  
compared to the case of corner-shared and edge-shared octahedrons. Hence, 
the importance of correlations from direct Coulomb interactions will be highly 
dilute. Also, the lack of orbital degeneracy will eliminate the ferro-orbital correlation 
and the associated C-type antiferromagnetic order.\cite{PhysRevLett.103.267001}

The strong covalency, absence of electronic correlations, lack of orbital degeneracy, 
and the dilute nature of the magnetic frustration in 
PdTe are in sharp contrast, in particular, to Fe chalcogenides, and in general, to Fe-based superconductors. 
Thus, PdTe will display dramatically weaker 
magnetic and superconducting tendencies than the Fe-based superconductors. 
One may conclude that the key ingredients for a high $T_c$ unconventional superconductivity 
are lost and pure PdTe may just be a low $T_c$ superconductor.
However, interesting magnetic correlation might be reactivated upon Fe doping.

\section{Summary}
We have performed self-consistent DFT and downfolding electronic band structure 
via symmetry respecting Wannier functions to study the electronic properties of PdTe. Our computations show that 
there is significant Pd-$d$ and Te-$p$ hybridization, larger crystal field splitting of Pd d-orbitals 
due to their near octahedral coordination, 
and higher filling of the Pd $d$-shell 
resulting in weaker magnetic frustration, strong covalency, 
and lack of orbital degeneracy, which quench  ferro-orbital correlations. 
The large $k_z$ variation of the Fermi surface topology also demonstrates a strong three-dimensional
character of the electronic structure. 
These features destroy the  ingredients needed for PdTe to be a high-T$_\mathrm{c}$ unconventional superconductor.
This case study provides a good contrast that highlights some important features 
(quasi-two-dimensionality, proximity to half-filling, weaker covalency, and higher orbital degeneracy) of 
Fe-based high-temperature superconductors.

\begin{acknowledgments}
\noindent We thank Carol Duran for carefully reading the manuscript. 
Work at LSU is funded by the National Science Foundation LA-SiGMA award: EPS 1003897. Work at 
BNL is supported by the U.S. Department of Energy (DOE) under contract DE-AC02-98CH10886.
Inter-institutional collaboration is supported by the DOE-CMCSN grant DE-AC02-98CH10886.
High performance computational resources are provided by the Louisiana Optical Network Initiative 
(LONI) and Brookhaven National Laboratory (BNL) clusters.  
\end{acknowledgments}


\end{document}